\documentclass{article}
\usepackage{authblk}
\usepackage{graphicx}
\usepackage[utf8]{inputenc}
\graphicspath{ {.home/user/Desktop/} }

\title{Online Information Search During COVID-19}
\author[1]{Benjamin Lucas}
\author[1]{Brendan Elliot}
\author[1]{Todd Landman}
\affil[1]{Rights Lab, University of Nottingham, United Kingdom}

\date{April 2020}
\begin{document}

  \maketitle
\section{Introduction}

Google Trends data has been previously employed in the study of public information search regarding everything from health conditions and medical symptoms (Cook et al., 2011; Dugas, et al., 2012 Cho et al., 2013) to socio-economic issue salience and impact (Vosen and Schmidt, 2011; Choi et al., 2012; Mellon, 2013, Preis et al., 2013). In general, such data affords researchers a perspective on what society does not know, or what society wants to find out prompted by or in response to, for example, developments in societal communication and news media events. In times of crisis, online public information search, via means such as Google searches, offers a window into the most urgent concerns of, and demands by society. By extension, such data offers a window into upcoming, and currently pressing demands on businesses, policymakers and researchers.
\hfill \break
\hfill \break
Presently, COVID-19 and the resulting Coronavirus pandemic (Ghinai  et al., 2020; Hellewell et al., 2020; Lillie et al., 2020; Huang et al., 2020; Petropoulos and Makridakis, 2020) represents one of the most substantial global threats to life and livelihood this century. Whilst frontline healthcare workers, government bodies and health researchers work in overdrive to solve this challenge, central (i.e. public health) and ancillary (e.g. social and economic) threats to society are revealed as the crisis unfolds. Policymakers and researchers thus cannot afford to lose sight of the impact of the COVID-19 crisis on all aspects of societies and economies, nor on other grand challenges facing humanity.
\hfill \break
\hfill \break
The purpose of this research note is twofold. First, we aim to highlight the potential of public-domain data sources to support rapidly mobilized public-domain data archiving during the COVID-19 crisis for future multi-disciplinary analysis.
\hfill \break
\hfill \break
Secondly, we aim to draw attention to existing related research with the aim of encouraging the rapid manoeuvre of the data science and computational social science communities, building on other timely works in the field (e.g. Basch et al., 2020; Cinelli et al., 2020; Ienca and Vayena, 2020; Ting et al., 2020), including in the use of Google Trends data (Hu et al., 2020; Husnayain et al., 2020; Lampos et al., 2020; Strzelecki 2020; Strzelecki and Rizun 2020).
\hfill \break
\hfill \break
We envisage this commentary having implications for research in fields such as (1) public health (e.g. epidemiological evidence from public web search behaviour), (2) economics and consumption (e.g. public behavior in relation to searches evidencing consumer interest in “N95 masks” in the present tense – through to broader consumer behavior shifts along the lines of ‘panic shopping’ and stockpiling, and demand spikes in retail, as well as business and labor market disruption broadly), with additional implications for (3) research into macro-level challenges facing humanity (e.g. along the lines of the UN SDGs – see: Sachs et al., 2019), and research into the reversion to broader societal ‘normality’ (The Guardian, 2020; MIT Technology Review, 2020).
\hfill \break
\hfill \break
An especially important consideration here is the extent to which current policy initiatives and policy instrument formulations pertaining to the broader betterment of society (e.g. fighting issues such as poverty, climate change, inequality and modern slavery), as well as those related to economic growth (in any form - including anything from infrastructure to innovation) (see: United Nations SDGs Knowledge Platform, 2020) may be disrupted - and how indicators and measures of issue salience and public behavior change (e.g. via information search) can serve as an early-warning system.
\hfill \break
\hfill \break
We wish in particular to underscore the importance of such data for, (4) future research examining the specifics of related searches (e.g. the extent to which different geographic locations and languages use specific search keywords, such as those related to COVID-19 symptoms, COVID-19 real-time update services, and various other health services, as well as other publicly salient topics), including their temporal-evolutionary nature. Finally, (5) the specifics of temporal dynamics of public information search during COVID-19 are also interesting from a plethora of angles - spanning focused investigations into specific temporally-defined gradients and ‘spikes’, through to forecasting.
\hfill \break

\begin{center}
   \includegraphics[width=120mm,scale=0.5]{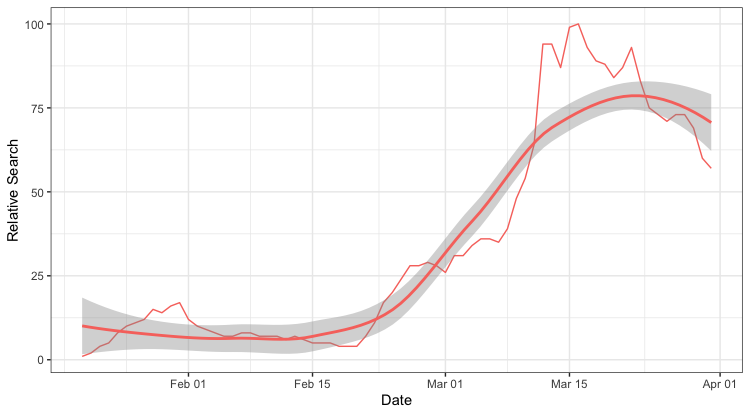}
   
   Figure 1: Global Google Search Activity for “coronavirus”
   
\end{center}

  \maketitle
\section{Approach}
Here, we provide a snapshot of public information search using Google Trends between the 1st of January 2020 and the 31st of March 2020. This data can be freely accessed using the public interface for Google Trends, or via packages such as ‘gtrendsR’ in R (Massicotte et al., 2016). We intend in particular here to illustratively flag developments in network-based approaches to the study of dynamic phenomena (e.g. Chi et al., 2010; Tabak et al., 2010; Garratt et al., 2014; Saba et al., 2014; Paci et al., 2020), given the use of such approaches in fields such as epidemiology and social influence studies. To this end, we focus on an extremely simple data collection and visualization framework.
\hfill \break
\hfill \break
First, we identified the global trend for Google searches using the keyword “coronavirus”. Next, we collected separate relative search volumes for each of the top countries searching this keyword (54 in total). We then limited the range further to only data corresponding with initial movement in the global trend – which meant trimming the time series to begin on the 20th of January 2020. This raw global trend is visualized in Figure 1. The y-axis shows relative search volume (normalized to between 0 and 100) for the search period. The x-axis shows the timeline of the focal period (see: Google Trends, 2020).
\hfill \break
\hfill \break

\hfill \break

\begin{center}
   \includegraphics[width=120mm,scale=0.5]{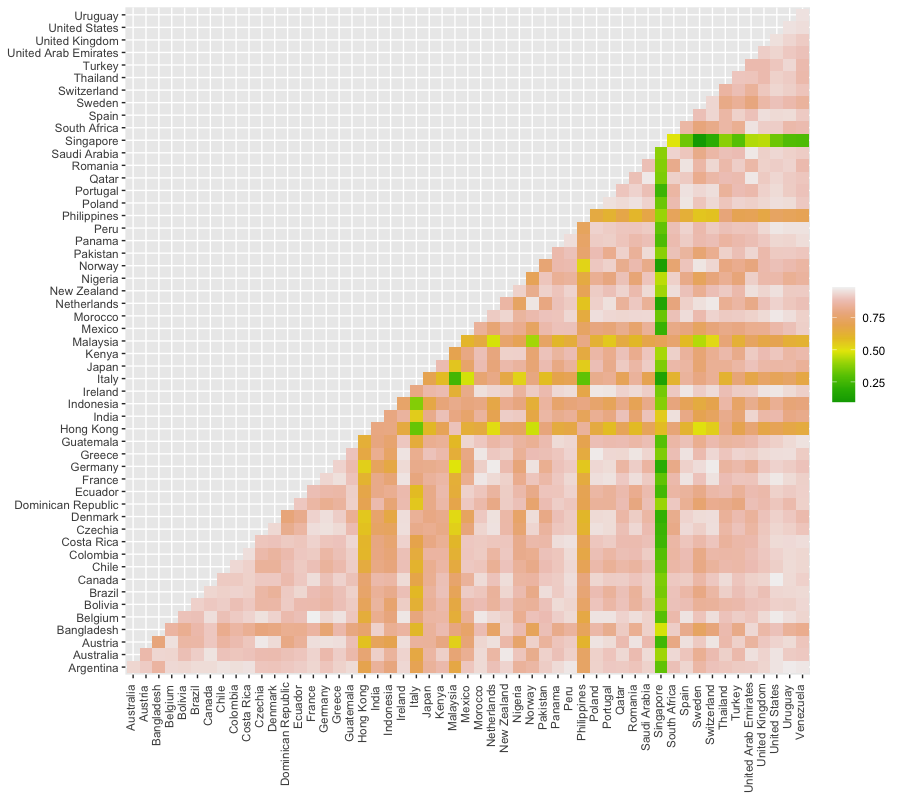}
   
   Figure 2: Correlations Between the 54 Highest Searching Locations
   
\end{center}

\hfill \break

To create a simple illustrative network example, we next calculated Spearman's rank correlations between all country-specific time series. This is visualized in Figure 2. This figure reveals that a majority of the time series are rather highly correlated, with searches in Singapore standing out as less correlated in particular.
\hfill \break
\hfill \break
To derive a clearer picture, We then used the resulting correlation matrix to calculate (a) a fully-connected weighted graph, then (b) a maximum-spanning tree (i.e. where only the strongest correlations between the 54 country nodes remain). This is visualized in Figure 3. Nodes are sized according to degree centrality.
 
\hfill \break

\begin{center}
   \includegraphics[width=130mm,scale=0.5]{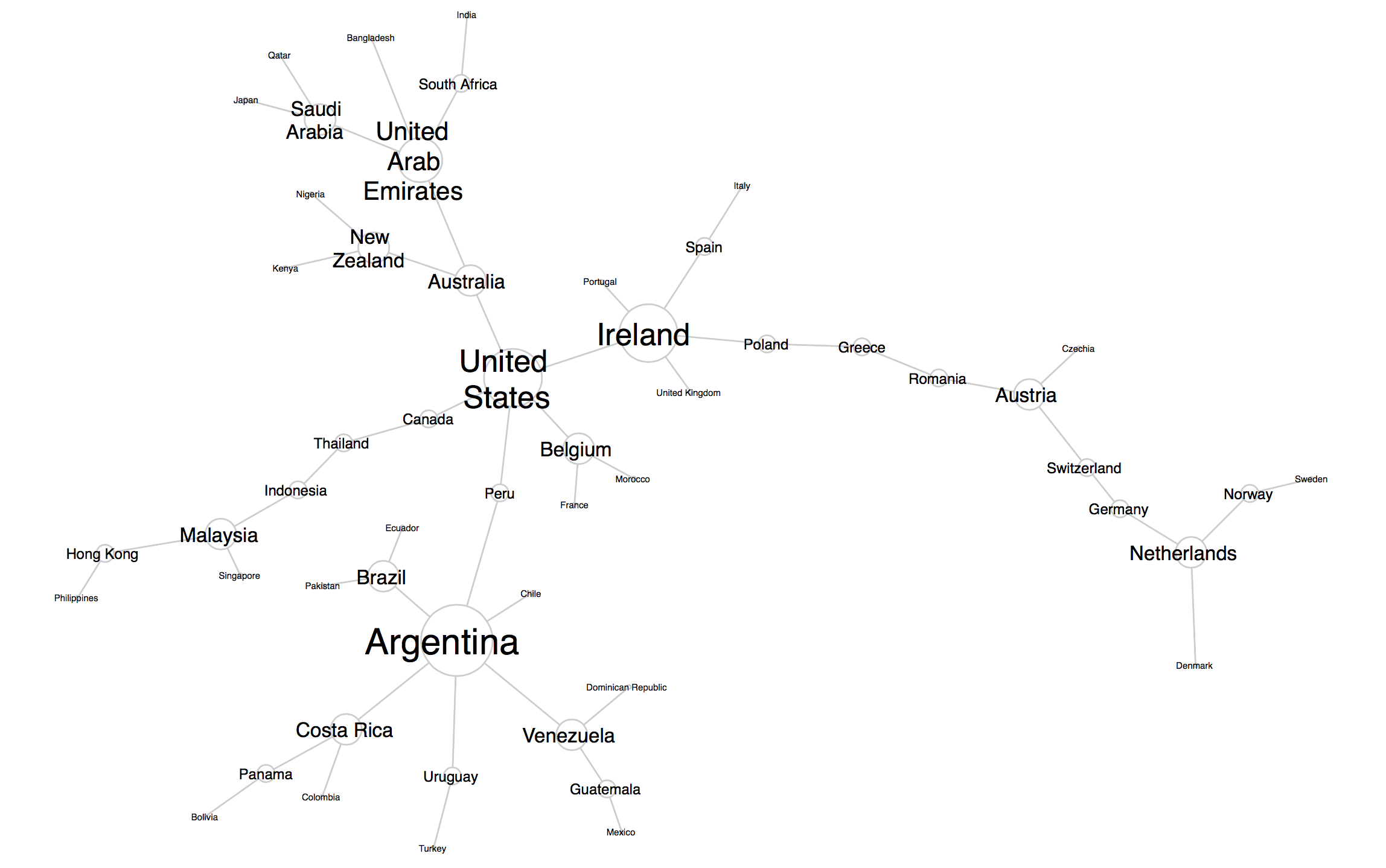}
\hfill \break
   Figure 3: Maximum Spanning Tree Based on Correlation
   
\end{center}

\hfill \break

Geographical clustering emerges based on visual inspection. The four largest branches of the maximum-spanning tree for instance reflect Asia, South America, Europe and Australasia and the Middle East.
\hfill \break
\hfill \break
For future research, data collection can be easily extended to accommodate lower search volume locations (as this research note includes only high-volume locations), and to include other languages (NB: “coronavirus” is a popular keyword at present reflecting the use of this spelling in the English language as well as many of the Latin script languages). The deliberate summarizing in this research note notwithstanding, we hope that interest in online public information search data is further spurred at this time.
\hfill \break
\hfill \break
For example, more in-depth analysis could be used to examine cointegration and contagion present within online information search in the context of COVID-19 (i.e. across geographies). To highlight the potential here, we provide a simple visual example in Figures 4 and 5, showing dominant search locations (for “coronavirus”) from the 1st of January 2020 to the 12th of April 2020. Different temporal resolutions are possible - but here, we show simple evidence of how public information search ‘lit up’ across the world at the beginning of the COVID-19 crisis.

\hfill \break
\hfill \break
Future research will also formally link public information search with a spectrum of exogenous factors - across news media activity, government (e.g. public health) communications, and user-driven momentum effects. As introduced, an especially interesting avenue for future research is linking public information search activity during the COVID-19 crisis with economic trends (e.g. such as unemployment - see: Fondeur and Karamé, 2013; Naccarato et al., 2018) in direct response to recent commentary from the United Nations (see: United Nations, 2020) and various recent news stories around the world related to labor market disruption, and disruption to commerce broadly (and government stimulus and contingency responses).

\begin{center}
   \includegraphics[width=150mm,scale=0.5]{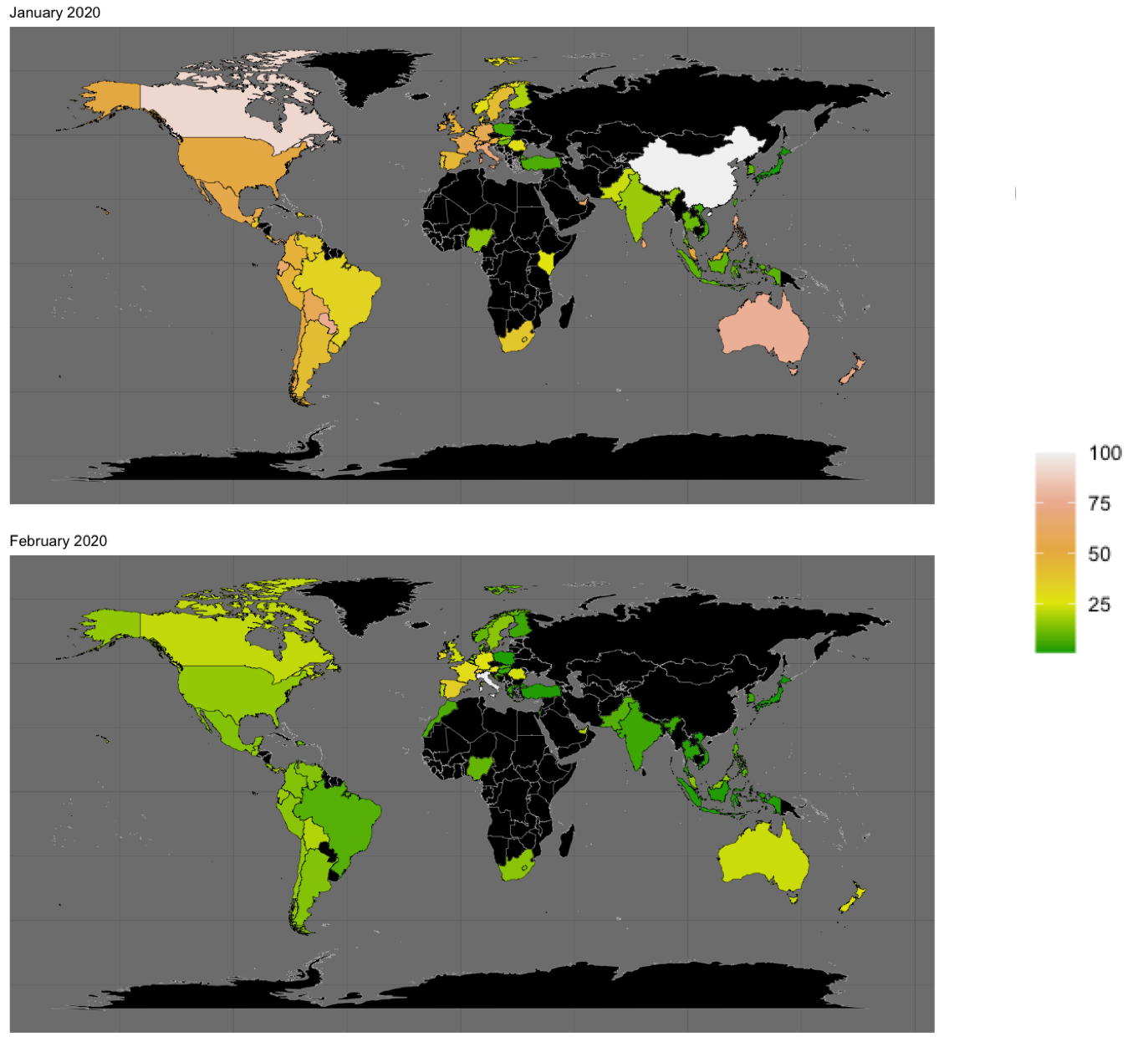}
   
   Figure 4: Choropleth of Global Search Activity for “coronavirus”
   
\end{center}

\begin{center}
   \includegraphics[width=150mm,scale=0.5]{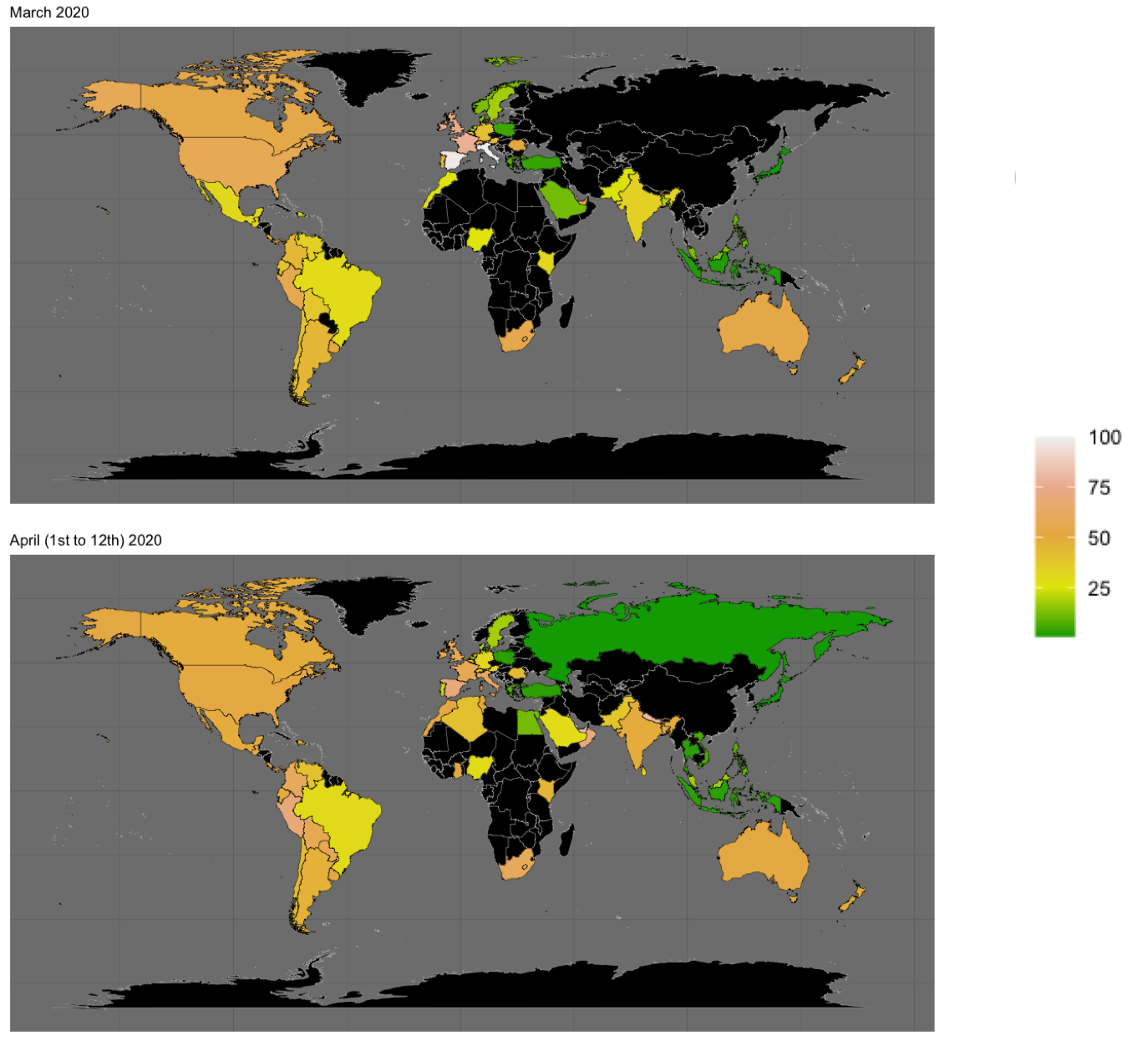}
   
   Figure 5: Choropleth of Global Search Activity for “coronavirus” (cont.)
   
\end{center}

\hfill \break
\hfill \break
\hfill \break
\hfill \break
\hfill \break
\hfill \break

  \maketitle
\section{Concluding Remarks}
Although it is difficult to rank societal priorities at times when the entire world’s population is facing catastrophe, situations of crisis do act as a catalyst for policymakers and researchers to think very comprehensively and efficiently about the full spectrum of humanity’s needs and the grand challenges within society that may threaten provisions for those needs, as well as about the available socio-technical apparatus to combat such threats.
\hfill \break
\hfill \break
Rapidly mobilized initiatives such as the UK Parliament’s “COVID-19 Out- break Expert Database” (UK Parliament, 2020) will likely reshape university technology transfer (see for background: Bercovitz and Feldman, 2006) for some time to come. Outside of health, epidemiology and virology research, domains such as computational social science and ubiquitous computing will host this - especially along the lines of frontline developments such as contagion tracing (e.g. Berke et al., 2020; Ferretti et al., 2020; Oliver et al., 2020) and symptom tracking (COVID Symptom Tracker, 2020), and additionally, given that opportunities to combine datastreams abound at this intersection (e.g. combining indicators from public web activity including information search with more robust measurement tools).
\hfill \break
\hfill \break
Encouragingly, many such advancements are already taking place, with examples including the “Oxford COVID-19 Government Response Tracker” (Hale et al., 2020) and the “Oxford COVID-19 Impact Monitor” (Qian and Saunders, 2020), and motivated further by projects such as recent work by Google to activate its mobility data resources in a privacy-sensitive way to support public health policy (Fitzpatrick and DeSalvo, 2020).
\hfill \break
\hfill \break
Importantly, an extensive range of topics within which social scientists specifically can play their part in fighting the crisis by supporting policymakers are emerging at present. These themes include “behavioral science”, “civil contingency planning and management”, “communicating uncertainty”, “consumer protection”, “crisis communications”, “health economics”, “human rights”, “industry”, “local government”, “public health”, and “public order” (UK Parliament, 2020). Google's own reporting (Google Trends, 2020) further highlights the possibilities here. The reporting includes a mixture of global trends along the lines of specific COVID-19 symptoms, COVID-19 testing, and a broad spectrum of related search queries and questions regarding economic impacts. It also includes results reflecting timely themes such as “social distancing” and “hand washing” - which involve both public health and sociological dimensions.
\hfill \break
\hfill \break
\hfill \break
\hfill \break
\hfill \break
\hfill \break
With Google searches acting as a primary portal for information search and gathering (and a major initial touchpoint or portal for service access) for a large percentage of the world’s population, perhaps the main short-term and long-term relevance here is around opening a discussion on information quality in the context of information search.
\hfill \break
\hfill \break
Questions around information quality have obvious implications for public information search regarding health conditions (see: Kitchens et al., 2014), but also more broadly for social order and well-being in terms of how the population evaluates everything from risks to actionable behavior change (e.g. in response to lock-downs). It may be for instance that search results can reveal saturation and critical-mass building along the lines of the effectiveness of certain policy measures, or that lagged effects in such data reveal important insights about the effectiveness of certain measures, as the public seeks information, builds an understanding, and acts accordingly.
\hfill \break
\hfill \break
This research note does not however represent an exhaustive assessment of public information search in the current COVID-19 crisis. For instance, other search engines in other markets, as well as differences between languages, and by extension, other keywords, need to be considered around the world. 
\hfill \break
\hfill \break
Specifically, “coronavirus” is an interesting keyword, as it is used in a number of English speaking countries, as well as in a number of other countries speaking languages of European origin. It is not by any means the sole terminology - not even within Europe. Beyond differences in vocabulary though, this is interesting from the viewpoint of investigating English and Latin script keyword usage for convenience, or other reasons (e.g. because of the influence of certain dominant communication channels and vehicles), in public information search around the world (see: Bokor, 2018). The diffusion of the COVID-19 naming convention (See: World Health Organization, 2020) is also in itself an interesting avenue for science communication research (e.g. Scheufele, 2014), although, at the time of writing, “coronavirus” remained the dominant search keyword globally by volume.
\hfill \break
\hfill \break
Against this backdrop, and in the continued interests of research efficiency and societal benefit creation - this research note thus encourages the continued pursuit of rapidly deployable approaches available to policymakers and researchers across all academic disciplines. We hope that this note encourages other researchers in social science and computational science to mobilize for the COVID-19 fight.

\hfill \break
\hfill \break
\hfill \break
\hfill \break
\hfill \break

\section{References}
Basch, C.H., Hillyer, G.C., Meleo-Erwin, Z.C., Jaime, C., Mohlman, J. and Basch, C.E., 2020. Preventive Behaviors Conveyed on YouTube to Mitigate Transmission of COVID-19: Cross-Sectional Study. JMIR Public Health and Surveillance, 6(2), p.e18807.
\hfill \break
\hfill \break
Bercovitz, J. and Feldman, M., 2006. Entrepreneurial universities and technology transfer: A conceptual framework for understanding knowledge-based economic development. The Journal of Technology Transfer, 31(1), pp.175-188.

\hfill \break
Berke, A., Bakker, M., Vepakomma, P., Raskar, R., Larson, K. and Pentland, A., 2020. Assessing Disease Exposure Risk With Location Histories And Protecting Privacy: A Cryptographic Approach In Response To A Global Pandemic. arXiv preprint arXiv:2003.14412.
\hfill \break
\hfill \break
Bokor, M.J., 2018. English dominance on the Internet. The TESOL encyclopedia of English language teaching, pp.1-6.
\hfill \break
\hfill \break
Chi, K.T., Liu, J. and Lau, F.C., 2010. A network perspective of the stock market. Journal of Empirical Finance, 17(4), pp.659-667.
\hfill \break
\hfill \break
Cho, S., Sohn, C.H., Jo, M.W., Shin, S.Y., Lee, J.H., Ryoo, S.M., Kim, W.Y. and Seo, D.W., 2013. Correlation between national influenza surveillance data and google trends in South Korea. PloS One, 8(12).
\hfill \break
\hfill \break
Choi, H. and Varian, H., 2012. Predicting the present with Google Trends. Economic Record, 88, pp.2-9.
\hfill \break
\hfill \break
Cinelli, M., Quattrociocchi, W., Galeazzi, A., Valensise, C.M., Brugnoli, E., Schmidt, A.L., Zola, P., Zollo, F. and Scala, A., 2020. The covid-19 social media infodemic. arXiv preprint arXiv:2003.05004.
\hfill \break
\hfill \break
Cook, S., Conrad, C., Fowlkes, A.L. and Mohebbi, M.H., 2011. Assessing Google flu trends performance in the United States during the 2009 influenza virus A (H1N1) pandemic. PloS One, 6(8).
\hfill \break
\hfill \break
COVID Symptom Tracker. 2020. Retrieved from: https://covid.joinzoe.com/
\hfill \break
\hfill \break
Dugas, A.F., Hsieh, Y.H., Levin, S.R., Pines, J.M., Mareiniss, D.P., Mohareb, A., Gaydos, C.A., Perl, T.M. and Rothman, R.E., 2012. Google Flu Trends: correlation with emergency department influenza rates and crowding metrics. Clinical Infectious Diseases, 54(4), pp.463-469
\hfill \break
\hfill \break
Ferretti, L., Wymant, C., Kendall, M., Zhao, L., Nurtay, A., Abeler-Dörner, L., Parker, M., Bonsall, D. and Fraser, C., 2020. Quantifying SARS-CoV-2 transmission suggests epidemic control with digital contact tracing. Science. Retrieved from: https://science.sciencemag.org/content/early/2020/04/09/science.abb6936.abstract
\hfill \break
Fitzpatrick. J. and DeSalvo., K., 2020. Helping public health officials combat COVID-19. Retrieved from: https://www.blog.google/technology/health/covid-19-community-mobility-reports
\hfill \break
\hfill \break
Fondeur, Y. and Karamé, F., 2013. Can Google data help predict French youth unemployment?. Economic Modelling, 30, pp.117-125.
\hfill \break
\hfill \break
Garratt, R.J., Mahadeva, L. and Svirydzenka, K., 2014. The great entanglement: The contagious capacity of the international banking network just before the 2008 crisis. Journal of Banking and Finance, 49, pp.367-385.
\hfill \break
\hfill \break
Ghinai, I., McPherson, T.D., Hunter, J.C., Kirking, H.L., Christiansen, D., Joshi, K., Rubin, R., Morales-Estrada, S., Black, S.R., Pacilli, M. and Fricchione, M.J., 2020. First known person-to-person transmission of severe acute respiratory syndrome coronavirus 2 (SARS-CoV-2) in the USA. The Lancet.
\hfill \break
\hfill \break
Google Trends. 2020. Coronavirus Search Trends. Retrieved from:https://trends.google.com/trends 
\hfill \break
The Guardian. 2020. ‘We can’t go back to normal’: how will coronavirus change the world?. Retrieved from: https://www.theguardian.com/world/2020/mar/31/how-will-the-world-emerge-from-the-coronavirus-crisis
\hfill \break
\hfill \break
Hale., T., Petherick., A., Phillips, T. and Webster. S., 2020. Variation in government responses to COVID-19. Blavatnik School of Government Working Paper Series – University of Oxford. Retrieved from: https://www.bsg.ox.ac.uk/research/publications/variation-government-responses-covid-19
\hfill \break
\hfill \break
Hellewell, J., Abbott, S., Gimma, A., Bosse, N.I., Jarvis, C.I., Russell, T.W., Munday, J.D., Kucharski, A.J., Edmunds, W.J., Sun, F. and Flasche, S., 2020. Feasibility of controlling COVID-19 outbreaks by isolation of cases and contacts. The Lancet Global Health.
\hfill \break
\hfill \break
Hu, D., Lou, X., Xu, Z., Meng, N., Xie, Q., Zhang, M., Zou, Y., Liu, J., Sun, G.P. and Wang, F., 2020. More Effective Strategies are Required to Strengthen Public Awareness of COVID-19: Evidence from Google Trends. Available at SSRN 3550008.
\hfill \break
\hfill \break
Huang, C., Wang, Y., Li, X., Ren, L., Zhao, J., Hu, Y., Zhang, L., Fan, G., Xu, J., Gu, X. and Cheng, Z., 2020. Clinical features of patients infected with 2019 novel coronavirus in Wuhan, China. The Lancet, 395(10223), pp.497-506.
\hfill \break
\hfill \break
Husnayain, A., Fuad, A. and Su, E.C.Y., 2020. Applications of google search trends for risk communication in infectious disease management: A case study of COVID-19 outbreak in Taiwan. International Journal of Infectious Diseases.
\hfill \break
\hfill \break
Ienca, M. and Vayena, E., 2020. On the responsible use of digital data to tackle the COVID-19 pandemic. Nature Medicine. https://doi.org/10.1038/s41591-020-0832-5
\hfill \break
\hfill \break
Kitchens, B., Harle, C.A. and Li, S., 2014. Quality of health-related online search results. Decision Support Systems, 57, pp.454-462.
\hfill \break
\hfill \break
Lampos, V., Moura, S., Yom-Tov, E., Cox, I.J., McKendry, R. and Edelstein, M., 2020. Tracking COVID-19 using online search. arXiv preprint arXiv:2003.08086.

\hfill \break
Lillie, P.J., Samson, A., Li, A., Adams, K., Capstick, R., Barlow, G.D., Easom, N., Hamilton, E., Moss, P.J., Evans, A. and Ivan, M., 2020. Novel coronavirus disease (Covid-19): the first two patients in the UK with person to person transmission. Journal of Infection.
\hfill \break
\hfill \break
Massicotte, P., Eddelbuettel, D. and Massicotte, M.P., 2016. Package ‘gtrendsR’.
\hfill \break
Mellon, J., 2013. Where and when can we use Google Trends to measure issue salience?. PS: Political Science and Politics, 46(2), pp.280-290.
\hfill \break
\hfill \break
MIT Technology Review. 2020. We’re not going back to normal. Retrieved from: https://www.technologyreview.com/s/615370/coronavirus-pandemic-social-distancing-18-months/
\hfill \break
\hfill \break
Naccarato, A., Falorsi, S., Loriga, S. and Pierini, A., 2018. Combining official and Google Trends data to forecast the Italian youth unemployment rate. Technological Forecasting and Social Change, 130, pp.114-122.
\hfill \break
\hfill \break
Oliver, N., Letouzé, E., Sterly, H., Delataille, S., De Nadai, M., Lepri, B., Lambiotte, R., Benjamins, R., Cattuto, C., Colizza, V. and de Cordes, N., 2020. Mobile phone data and COVID-19: Missing an opportunity?. arXiv preprint arXiv:2003.12347.
\hfill \break
\hfill \break
Paci, P., Fiscon, G., Conte, F., Licursi, V., Morrow, J., Hersh, C., Cho, M., Castaldi, P., Glass, K., Silverman, E.K. and Farina, L., 2020. Integrated transcriptomic correlation network analysis identifies COPD molecular determinants. Scientific Reports, 10(1), pp.1-18.
\hfill \break
\hfill \break
Preis, T., Moat, H.S. and Stanley, H.E., 2013. Quantifying trading behavior in financial markets using Google Trends. Scientific Reports, 3, p.1684.
\hfill \break
\hfill \break
Qian, M. and Saunders., A., 2020. Oxford COVID-19 Impact Monitor. Retrieved from: https://oxford-covid-19.com/
\hfill \break
\hfill \break
Saba, H., Vale, V.C., Moret, M.A. and Miranda, J.G.V., 2014. Spatio-temporal correlation networks of dengue in the state of Bahia. BMC public health, 14(1), p.1085.
\hfill \break
\hfill \break
Sachs, J.D., Schmidt-Traub, G., Mazzucato, M., Messner, D., Nakicenovic, N. and Rockström, J., 2019. Six transformations to achieve the sustainable development goals. Nature Sustainability, 2(9), pp.805-814.
\hfill \break
\hfill \break
Scheufele, D.A., 2014. Science communication as political communication. Proceedings of the National Academy of Sciences, 111(Supplement 4), pp.13585-13592.
\hfill \break
\hfill \break
Strzelecki, A., 2020. The Second Worldwide Wave of Interest in Coronavirus since the COVID-19 Outbreaks in South Korea, Italy and Iran: A Google Trends Study. arXiv preprint arXiv:2003.10998.
\hfill \break
\hfill \break
Strzelecki, A. and Rizun, M., 2020. Infodemiological study using google trends on coronavirus epidemic in Wuhan, China. arXiv preprint arXiv:2001.11021.
\hfill \break
\hfill \break
Tabak, B.M., Serra, T.R. and Cajueiro, D.O., 2010. Topological properties of stock market networks: The case of Brazil. Physica A: Statistical Mechanics and its Applications, 389(16), pp.3240-3249.
\hfill \break
\hfill \break
Ting, D.S.W., Carin, L., Dzau, V. and Tien Y. Wong. 2020. Digital technology and COVID-19. Nature Medicine. https://doi.org/10.1038/s41591-020-0824-5
\hfill \break
\hfill \break
UK Parliament. 2020. COVID-19 Outbreak Expert Database. Retrieved from: https://www.parliament.uk/covid19-expert-database
\hfill \break
\hfill \break
United Nations. 2020. COVID-19: impact could see 195 million job losses, says ILO chief. Retrieved from: https://www.un.org/sustainabledevelopment/blog/2020/04/covid-19-impact-could-see-195-million-job-losses-says-ilo-chief/
\hfill \break
\hfill \break
United Nations SDGs Knowledge Platform, 2020. Sustainable Development Goals. Retrieved from: https://sustainabledevelopment.un.org
\hfill \break
\hfill \break
Vosen, S. and Schmidt, T., 2011. Forecasting private consumption: survey‐based indicators vs. Google trends. Journal of Forecasting, 30(6), pp.565-578.
\hfill \break
\hfill \break
World Health Organization. 2020. Naming the coronavirus disease (COVID-19) and the virus that causes it. Retrieved from: https://www.who.int/emergencies/diseases/novel-coronavirus-2019/technical-guidance/naming-the-coronavirus-disease-(covid-2019)-and-the-virus-that-causes-it
\end{document}